\documentclass[12pt]{iopart}
\usepackage{iopams} 
\usepackage{graphicx}
\usepackage{braket}
\usepackage{xspace}

\newcommand{\EF}{$E_\mathrm{F}$\xspace}
\newcommand{\Cef}{$\mathrm{Ce}~4f$\xspace}
\newcommand{\Ced}{$\mathrm{Ce}~3d$\xspace}
\newcommand{\Cedf}{$\mathrm{Ce}~3d-4f$\xspace}
\newcommand{\Ird}{$\mathrm{Ir}~5d$\xspace}
\newcommand{\orb}[2]{$\mathrm{ #1 } ~ #2 $\xspace}
\newcommand{\hn}[1]{$h\nu #1~\mathrm{eV}$\xspace}
\newcommand{\EB}[1]{$E_{\mathrm{B}} #1~\mathrm{eV}$\xspace}
\newcommand{\Tc}[1]{$T_{\mathrm{SC}}= #1~\mathrm{K}$\xspace}
\newcommand{\nf}[1]{$n_{4f} #1 $\xspace}
\newcommand{\CeIr}{$\mathrm{CeIr_3}$\xspace}
\newcommand{\LaIr}{$\mathrm{LaIr_3}$\xspace}
\newcommand{\ThIr}{$\mathrm{ThIr_3}$\xspace}

\begin{document}

\title[{\small Impact of the $\mathrm{Ce}~4f$ states in the electronic structure of the intermediate-valence superconductor CeIr$_3$}]{Impact of the $\mathrm{Ce}~4f$ states in the electronic structure of the intermediate-valence superconductor CeIr$_3$}

\author{
Shin-ichi~Fujimori$^1$,
Ikuto~Kawasaki$^1$,
Yukiharu~Takeda$^1$,
Hiroshi~Yamagami$^{1,2}$,
Norimasa~Sasabe$^3$,
Yoshiki~J.~Sato$^{4, 5}$,
Ai~Nakamura$^4$,
Yusei~Shimizu$^4$,
Arvind~Maurya$^{4, 6}$,
Yoshiya~Homma$^4$,
Dexin~Li$^4$,
Fuminori~Honda$^{4, 7}$,
and Dai~Aoki$^4$
}

\address{$^1$ Materials Sciences Research Center, Japan Atomic Energy Agency, Sayo, Hyogo 679-5148, Japan}
\address{$^2$ Department of Physics, Faculty of Science, Kyoto Sangyo University, Kyoto 603-8555, Japan}
\address{$^3$ Japan Synchrotron Radiation Research Institute, SPring-8, Sayo, Hyogo 679-5148, Japan}
\address{$^4$ Institute for Materials Research, Tohoku University, Oarai, Ibaraki 311-1313, Japan}
\address{$^5$ Department of Physics, Faculty of Science and Technology, Tokyo University of Science, Noda, Chiba 278-8510, Japan}
\address{$^6$ Department of Physics, School of Physical Sciences, Mizoram University, Aizawl 796 004, India}
\address{$^7$ Central Institute of Radioisotope Science and Safety Management, Kyushu University, Motooka 744, Fukuoka Nishi, Fukuoka 819-0395, Japan}

\ead{fujimori@spring8.or.jp}
\vspace{10pt}
\begin{indented}
\item[] \today
\end{indented}

\begin{abstract}
The electronic structure of the $f$-based superconductor $\mathrm{CeIr_3}$ was studied by photoelectron spectroscopy.
The energy distribution of the $\mathrm{Ce}~4f$ states were revealed by the $\mathrm{Ce}~3d-4f$ resonant photoelectron spectroscopy.
The $\mathrm{Ce}~4f$ states were mostly distributed in the vicinity of the Fermi energy, suggesting the itinerant character of the $\mathrm{Ce}~4f$ states.
The contribution of the $\mathrm{Ce}~4f$ states to the density of states (DOS) at the Fermi energy was estimated to be nearly half of that of the $\mathrm{Ir}~5d$ states, implying that the $\mathrm{Ce}~4f$ states have a considerable contribution to the DOS at the Fermi energy.
The $\mathrm{Ce}~3d$ core-level and $\mathrm{Ce}~3d$ X-ray absorption spectra were analyzed based on a single-impurity Anderson model.
The number of the $\mathrm{Ce}~4f$ states in the ground state was estimated to be $0.8-0.9$, which is much larger than the values obtained in the previous studies (i.e., $0-0.4$).
\end{abstract}

\submitto{Electronic Structure}

\maketitle

\section{Introduction}
\CeIr is a $f$-based superconductor with a relatively high transition temperature (\Tc{3.1}), and its superconducting properties have attracted much attention in recent years \cite{CeIr3_Sato1, CeIr3_XPS, CeIr3_Sato2, CeIr3_Adroja}.
Its thermodynamical properties and $\mu$SR measurements suggest \CeIr is a weak-coupling superconductor with an anisotropic $s$-wave gap \cite{CeIr3_Sato1,CeIr3_Adroja}. 
One of the key issues of this compound is the role of the \Cef states in superconductivity.
Isostructural analogue compounds \LaIr (\Tc{2.5} \cite{LaIr3_1,LaIr3_2}) and \ThIr (\Tc{4.41} \cite{ThIr3}) also exhibit the superconductivity with similar or higher transition temperatures; hence, the role of the \Cef states in the superconductivity of \CeIr is considered to be negligible \cite{XIr3}.
Experimentally, the \Cef states in \CeIr are suggested as an intermediate-valence state, but the occupancy of the $4f$ states \nf{} is a matter of controversies.
Based on the relationship between the transition temperature and the valence states in $\mathrm{(La,Th)Ir_3}$ alloys, $n_{4f}$ is estimated as \nf{\sim 0.4} \cite{CeIr3_CeRu2}.
G\'{o}rnicka \etal recently suggested that \nf{\sim 0} based on the weak-temperature-dependent magnetic susceptibility of \CeIr. They argued that the contributions from the \Cef states at the Fermi energy are negligible in this compound \cite{CeIr3_XPS}.
On the contrary, the experimental specific heat coefficient ($\gamma_{\mathrm{exp}} = 25.1~\mathrm{mJ/mol \cdot K^2}$) is much larger than that obtained by the GGA calculation ($\gamma_{\mathrm{GGA}} = 10.2~\mathrm{mJ/mol \cdot K^2}$) \cite{CeIr3_XPS}, suggesting that there should be a considerable contribution from the correlated \Cef states at the Fermi energy.
Gutowska \etal recently calculated the electronic structure of \CeIr based on the dynamical mean field theory (DMFT) and reproduced the experimental specific heat coefficient and the electron--phonon coupling \cite{CeIr3_DMFT}.
Their calculation showed that  the \Cef states had a finite contribution at the Fermi energy, and its occupation number was \nf{\sim 0.8}. 
Thus, the nature of the \Cef states of this compound is controversial.

In this study, we have investigated the \Cef states of \CeIr by the \Cedf resonant photoelectron spectroscopy (RPES), X-ray absorption spectroscopy (XAS), and core-level photoelectron spectroscopy.
Accordingly, the RPES study shows that the \Cef states form a sharp peak at the Fermi energy, suggesting that the \Cef states in \CeIr have a considerable contribution to the Fermi energy despite the stronger contribution from the \Ird states.
The results of the analysis of the \Ced XAS and core-level spectra based on the single-impurity Anderson model reveal that \nf{= 0.8-0.9}, contrast to the predictions of the previous studies \cite{CeIr3_XPS, CeIr3_CeRu2}.
These results argue that the \Cef states are in an intermediate-valence state, and the contribution from the \Cef states to superconductivity is not negligible in \CeIr.

\section{Experimental Procedure}
Single crystals of \CeIr were prepared using the Czochralski method in a tetra arc furnace under Ar atmosphere.
The details of the sample growth and their characterizations were presented in Refs.~\cite{CeIr3_Sato1, CeIr3_Sato2, CeIr3_Sato3}.
Photoemission experiments were performed at the soft X-ray beamline BL23SU of SPring-8 \cite{BL23SU2}.
The total energy resolution at the on-resonant condition (\hn{\sim 881}) was approximately $65~\mathrm{meV}$, while that in the \orb{Ce}{3d} core-level measurements was approximately $190~\mathrm{meV}$.
Clean sample surfaces were obtained by cleaving high-quality single crystals {\it in situ} under an ultra-high vacuum condition.
The sample temperature was kept at $20~\mathrm{K}$ during the measurements.

\section{Results and Discussion}
\begin{figure}
	\centering
	\includegraphics[scale=0.5]{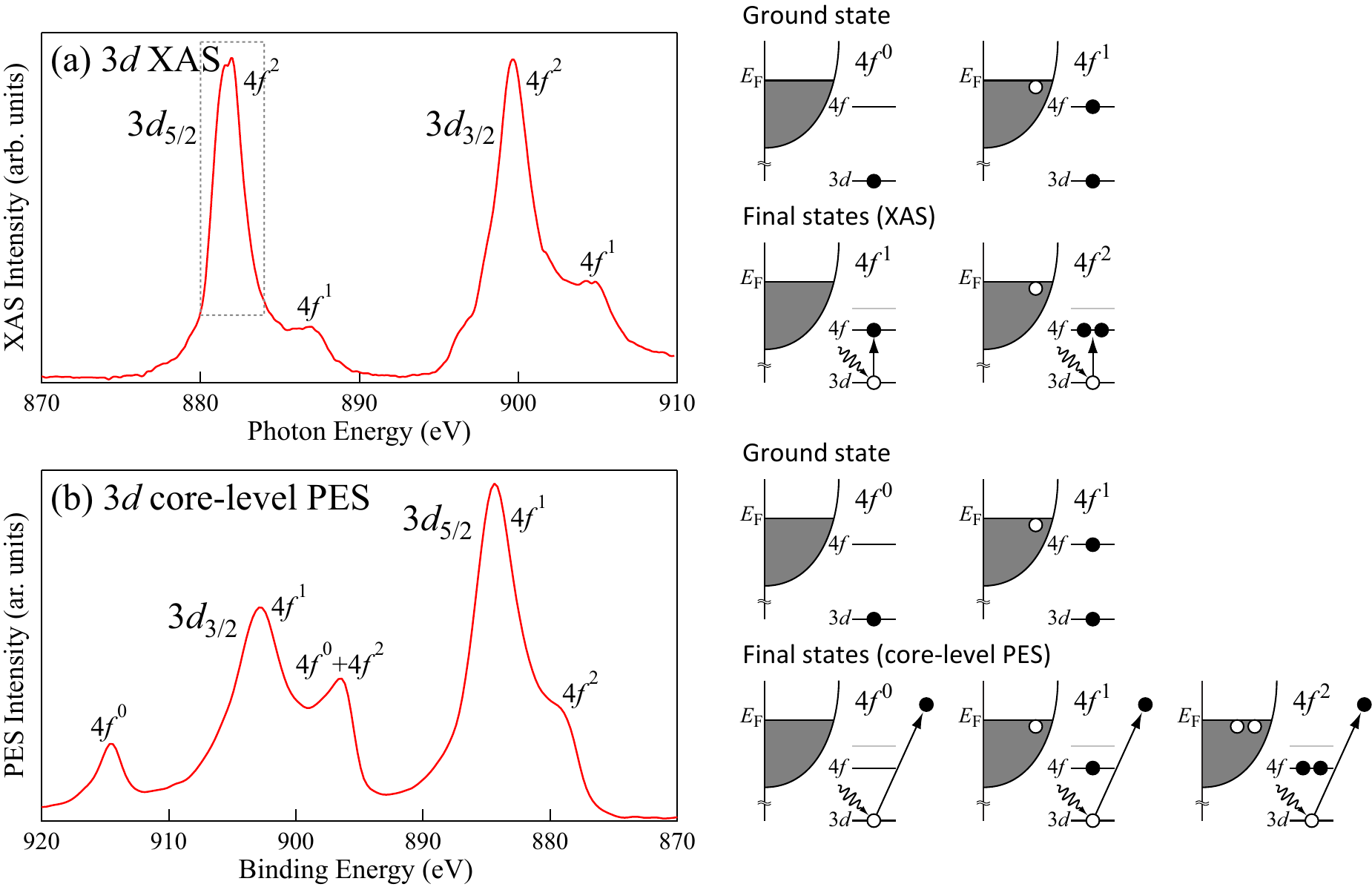}
	\caption{\Ced XAS and core-level spectra of \CeIr.
The schematic figures of the ground and final states in the XAS and core-level spectra are shown on the right.
(a) \Ced XAS spectrum of \CeIr consisting of $3d_{5/2}$ and $3d_{3/2}$ components.
Each has $4f^1$ and $4f^2$ final-state peaks.
The dashed square corresponds to the region shown in Fig.~\ref{RPES}(a).
(b) \Ced core-level spectrum of \CeIr consisting of $3d_{5/2}$ and $3d_{3/2}$ components.
Each has $4f^0$, $4f^1$, and $4f^2$ final-state peaks.
}
	\label{XAS_XPS}
\end{figure}
Figure~\ref{XAS_XPS} shows the \Ced XAS and core-level spectra of \CeIr together with the schematic figures of their ground and final states in the XAS and core-level spectra.
Figure~\ref{XAS_XPS} (a) depicts the \Ced XAS spectrum of \CeIr consisting of $3d_{5/2}$ and $3d_{3/2}$ components.
Each had $4f^1$ and $4f^2$ final-state peaks.
The $4f^1$ and $4f^2$ peaks originated from the transitions from the $4f^0$ and $4f^1$ configurations in the ground state, respectively.
The existence of these two peaks indicated that the \Cef states in \CeIr are in an intermediate-valence state.
Figure~\ref{XAS_XPS}(b) shows the \Ced core-level spectrum of \CeIr.
The photon energy was \hn{=1200}. 
The spectrum was essentially consistent with that in a previous XPS study \cite{CeIr3_XPS}; however, each peak was well resolved by the better energy resolution in the present experiment.
The $4f^0$ peak originates from the $4f^0$ configuration in the ground state.
The enhanced intensity of the $4f^0$ peak again suggested that the \Cef states were in a strong mixed valence state.
The calculation based on the single-impurity Anderson model suggested that $1-n_{4f} \approx I(f^0)/[I(f^0)+I(f^1)+I(f^2)]$, where $I(f^n)$ is the $4f^n$ final-state peak intensity \cite{GS}.
Thus, \nf{} of \CeIr should be close to, but less than unity in the present case.
The $4f^2$ final-state peak intensity was enhanced in the spectrum and was much stronger than in the case of heavy fermion compounds $\mathrm{Ce}T\mathrm{In}_5 (T=\mathrm{Co}, \mathrm{Rh},\mathrm{Ir})$ as an example \cite{Ce115XPS}.
The $4f^2$ final-state peak intensity is more enhanced in a more hybridized compound \cite{Ce_SIAM}; hence, the enhanced intensity in \CeIr indicates the existence of a strong $f-d$ hybridization in \CeIr.
The later part of this section will present a detailed analysis of these spectra.

\begin{figure}
	\centering 
	\includegraphics[scale=0.5]{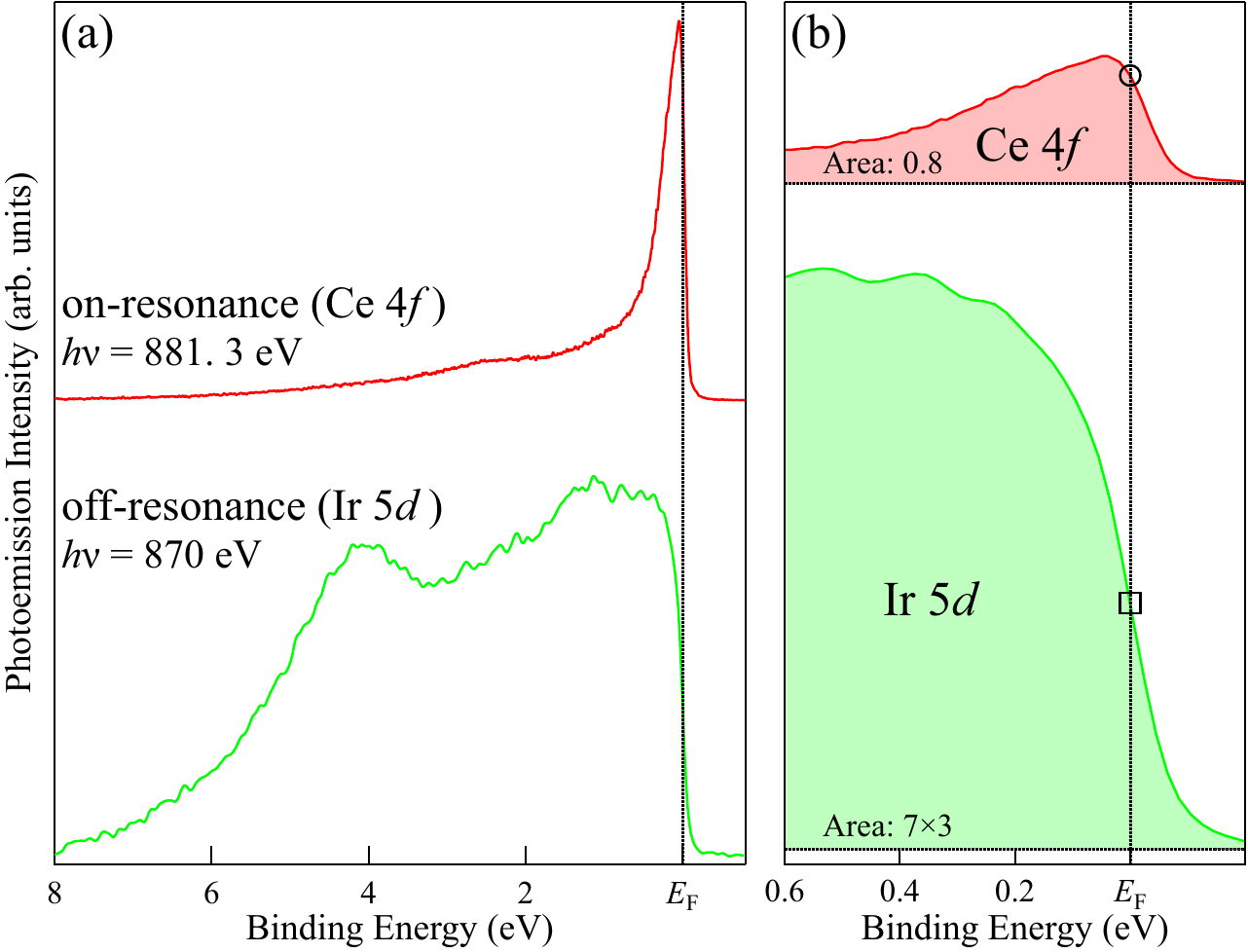}
	\caption{Resonant photoemission spectra of \CeIr.
	(a) On-resonance and off-resonance photoemission spectra of \CeIr measured at $h \nu=881.3$ and $870~\mathrm{eV}$, which reflect the contributions mainly from the \Cef and \Ird states, respectively.
	(b) On- and off-resonance spectra in the very vicinity of \EF.
	The areas of the spectra are normalized to the nominal numbers of electrons of the \Cef and \Ird states ($n_{4f}:n_{5d}=0.8:7 \times 3$).	
	}
	\label{RPES}
\end{figure}
Figure~\ref{RPES} summarizes the RPES measurement of \CeIr.
Figure~\ref{RPES} (a) shows the on- and off-resonance photoemission spectra of \CeIr measured at \hn{=881.3} and \hn{=870}, which reflects contributions mainly from the \Cef and \Ird states, respectively \cite{atomic}.
The \Cef states form a very sharp peak at the Fermi energy, indicating its considerable contribution to this energy. 
Meanwhile, a weak hump structure was found to exist around \EB{=2.3}.
This was assigned to the contribution from the $4f^0$ configuration in the final state.
In contrast to the \Cef states, the \Ird states form a broad band structure distributed in a wide energy range, with an overall spectral shape consistent with the GGA \cite{CeIr3_XPS} and DFT + DMFT calculations \cite{CeIr3_DMFT}.
We attempted herein to estimate the practical contributions from each orbital at the Fermi energy. 
Figure~\ref{RPES} (b) depicts the on- and off-resonance spectra in the very vicinity of the Fermi energy.
The areas of the spectra were normalized to the nominal numbers of the electrons of the \Cef and \Ird states in \CeIr ($n_{4f}\!:\!n_{5d}=0.8\!:\!7 \times 3$).
The \Cef and the \Ird spectra intensities can be directly compared. 
In practice, the occupation number of the \Cef states in \CeIr is estimated to be approximately 0.8 as we shall discuss later.
Both the \Ird and \Cef states have considerable contributions to the Fermi energy.
The contribution from the \Cef states at the Fermi energy was approximately 44 \% of that from the \Ird states; thus, the contribution from the \Cef states at the Fermi energy was estimated to be approximately 31 \% of the total DOS.
Note that the \Cef states could form a very narrow peak at the Fermi energy, but might be smeared out by the present energy resolution ($65~\mathrm{meV}$).
In this case, the substantial contribution from the \Cef states may be higher than this estimation.

A further interesting point was that we found no trace of the spin-orbit-coupled $4f_{7/2}$ multiplet, located at $E_{\mathrm{B}} \sim 280~\mathrm{meV}$. 
The $4f_{7/2}$ multiplet originates from the final state effect, and its intensity becomes weaker than that of the peak at \EF in the Ce compound with higher Kondo temperature ($T_{\mathrm{K}}$) \cite{Ce4f_SO}.
Thus, the existence of the sharp peak at \EF and the absence of the $4f_{7/2}$ multiplet peak in the \Cef spectrum indicate that $T_{\mathrm{K}}$ is very high in this compound.
The spectral shape of the \Cef spectrum is very similar not with the GGA calculation \cite{CeIr3_XPS}, but with the DMFT calculation \cite{CeIr3_DMFT}.
This is consistent with the itinerant but correlated nature of the \Cef states in \CeIr.

We tried qualitatively understand the $\mathrm{Ce}~3d$ core-level and XAS spectra by analyzing them based on the single-impurity Anderson model (SIAM) \cite{SIAM1,SIAM2,SIAM3}. 
The Hamiltonian of the system is given as

\begin{equation}
	H=H_1+H_2
\end{equation}

\noindent
where

\begin{eqnarray}
		H_1 & = \sum_{k,\nu} \varepsilon_v (k) a^{\dagger}_v (k, \nu) a_v (k, \nu)
			+ \varepsilon^0_f \sum_{\nu} {a_f}^{\dagger} (\nu) a_f (\nu) \nonumber \\
		&	+ \varepsilon_d \sum_{\mu} {a_d}^{\dagger} (\mu) a_d (\mu)  
			+ U_{ff} \sum_{\nu > \nu'} {a_f}^{\dagger} (\nu) a_f (\nu) {a_f}^{\dagger} (\nu') a_f (\nu') \nonumber \\
		&	+ \frac{V}{\sqrt{N}} \sum_{k, \nu} [ {a_v}^{\dagger} (k, \nu) a_f (\nu) + {a_f}^{\dagger} (\nu) a_v (k, \nu) ]  \nonumber \\
		&	-U_{fc} \sum_{\mu, \nu} a_d (\mu)^{\dagger} {a_d} (\mu) a_f^{\dagger} (\nu) a_f (\nu)
\end{eqnarray}

\noindent
and

\begin{eqnarray}
		H_2  & = \frac{1}{2} \sum_{\nu_1, \nu_2, \nu_3, \nu_4} g_{ff} (\nu_1, \nu_2, \nu_3, \nu_4)
		{a_f}^{\dagger} (\nu_1) {a_f}^{\dagger} (\nu_2) a_f (\nu_3) a_f (\nu_4) \nonumber \\
		& + \sum_{\nu_1, \nu_2, \mu_1, \mu_2} g_{fd} (\nu_1, \nu_2, \mu_1, \mu_2)
		{a_f}^{\dagger} (\nu_1) {a_d}^{\dagger} (\mu_1) a_d (\mu_2) a_f (\nu_2) \nonumber \\
		& + \zeta_d \sum_{\mu_1, \mu_2} \alpha_{\mu_1, \mu_2} {a_d}^\dagger (\mu_1) a_d (\mu_2)
		+ \zeta_f \sum_{\nu_1, \nu_2} \beta_{\nu_1, \nu_2} {a_f}^\dagger (\nu_1) a_f (\nu_2)
\end{eqnarray}

\noindent
$H_1$ describes the SIAM with the valence band level $\varepsilon_v (k)$, $4f$ level $\varepsilon^0_f$ and $3d$ level $\varepsilon_d$.
$ {a_v}^\dagger (k,\nu) $, $ {a_f}^\dagger (\nu) $, and ${a_d}^\dagger (\mu)$ are the electron creation operators in these states.
$k$ denotes the energy level index ($k=1, 2, \cdots, N$) in the valence band.
$\nu$ and $\mu$ denote the combined indices to specify both the spin and orbital degeneracies of the $f$ ($\nu =1, 2, \cdots, N_f$) and $d$ ($\mu =1, 2, \cdots, N_d$) states with the numbers of the degeneracy of $N_f=14$ and $N_d=10$, respectively.
$H_2$ represents the multiplet coupling effect described by Slater integrals $F^2$, $F^4$, and $F^6$ between $4f$ electrons ($g_{ff}$), the Slater integrals $F^2$, $F^4$, $G^1$, $G^3$, and $G^5$ between the $4f$ electrons and the $3d$ core hole ($g_{fd}$), and the spin-orbit interactions of the $3d$ states ($\zeta_d$) and the $4f$ states ($\zeta_f$).
The Slater integrals were reduced down to 80\% of its atomic Hartree--Fock values, which are calculated by using the Cowan program \cite{Cowan}.
The valence band energies $\varepsilon_v (k)$ were taken as

\begin{equation}
	\varepsilon_v (k) = \varepsilon^0_v - \frac{W}{2} + \frac{W}{N} ( k-\frac{1}{2} ), \;
	k = 1, 2, \cdots N,
\end{equation}

\noindent
where $\varepsilon^0_v$ and $W$ are the center and width of the valence band, respectively.
The $N$ and $W$ values were $N=20$ and $W=5~\mathrm{eV}$, respectively.
The $3d$ core-level ($F_{3d~\mathrm{XPS}} (\omega)$) and $3d$ XAS ($F_{3d~\mathrm{XAS}} (\Omega)$) spectral functions are given as follows:

\begin{equation}
	F_{3d~\mathrm{XPS}} (\omega) = \sum_{f,\mu} |\braket{f|a_d (\mu)|g}|^2 \frac{\Gamma/\pi}{(E_g+\omega-E_f)^2+\Gamma^2}
	\label{F_XPS}
\end{equation}

\begin{equation}
	F_{3d~\mathrm{XAS}} (\Omega) = \sum_{f} |\braket{f|T_1|g}|^2 \frac{\Gamma/\pi}{(E_g+\Omega-E_f)^2+\Gamma^2}
	\label{F_XAS}
\end{equation}

\noindent
where $\omega$, $\Omega$, and $T_1$ are the binding energy of the photoelectrons, photon energy, and optical dipole transition from the $3d$ to $4f$ states, respectively.
Note that $\ket{f}$ in Eqs.~(\ref{F_XPS}) and (\ref{F_XAS}) represent $N-1$ and $N$ electron states, respectively 

\begin{figure*}
	\centering
	\includegraphics[scale=0.5]{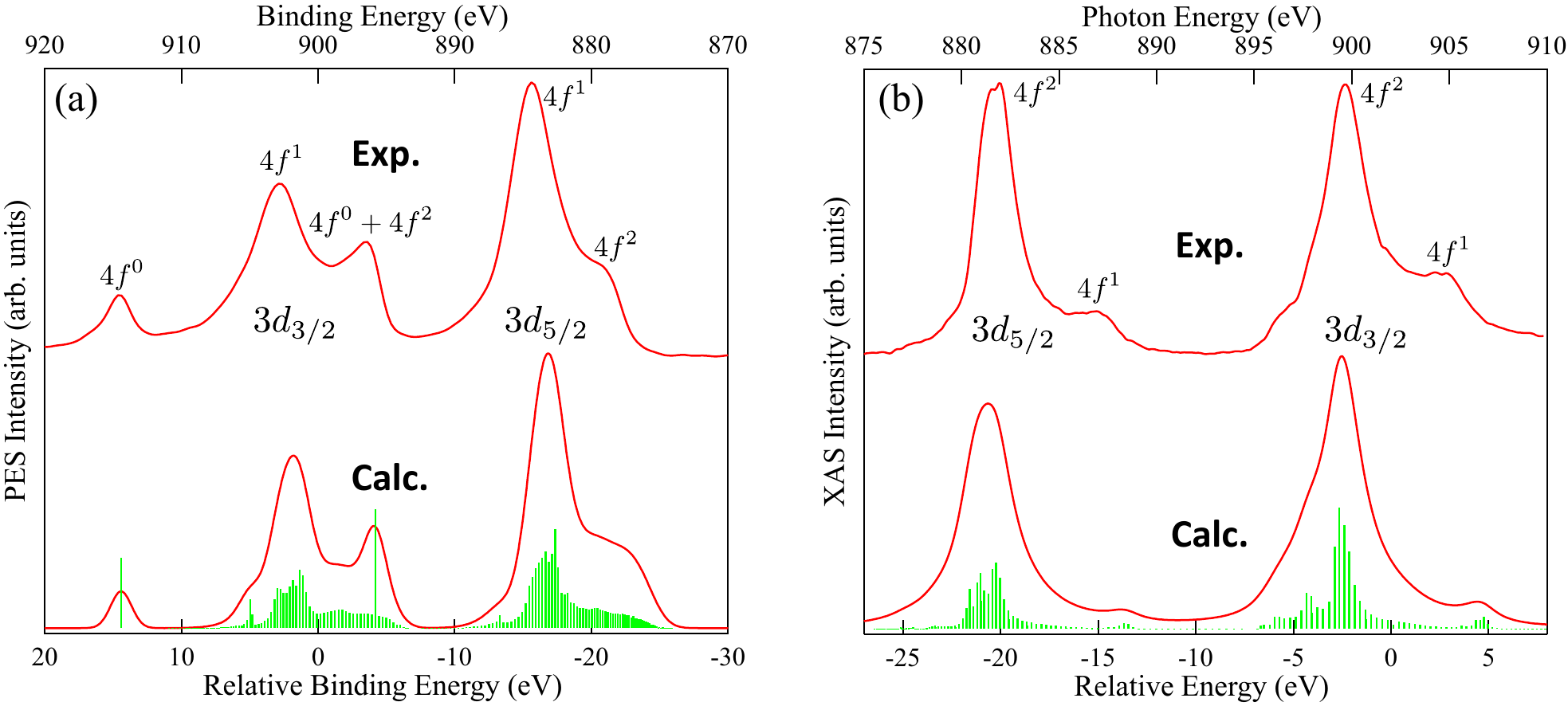}
	\caption{(a) \Ced XAS and (b) core-level spectra of \CeIr and their analysis results based on the SIAM.
	The histogram represents the matrix elements in the calculation.
	The curves depict the spectra with lifetime broadening represented in Eqs. (5) and (6).
}
	\label{Ce3d}
\end{figure*}
Figure~\ref{Ce3d} shows the analysis result of the (a) \Ced core-level and (b) XAS spectra.
We used the following parameters here: $E_{\mathrm{F}} - \varepsilon^0_f=-2.3~\mathrm{eV}$; $U_{ff}=7.0~\mathrm{eV}$; $U_{fc}=10.5~\mathrm{eV}$; and $V=0.5~\mathrm{eV}$.
The ratios of the $4f^0$, $4f^1$, and $4f^2$ configurations in the ground state were 0.111, 0.812, and 0.077, respectively.
The occupation number of the \Cef states in the ground state was \nf{=0.966}.
Figure~\ref{Ce3d} (a) compares the experimental \Ced core-level spectrum of \CeIr and the analysis result.
The histogram represents the matrix elements in the calculation.
The curves depict the spectra with the life--time broadening represented in Eqs. (5) and (6).
The experimental spectrum was reproduced well by the SIAM using these parameters.
Figure ~\ref{Ce3d}(b) displays the result of the \Ced XAS spectrum calculated by the same parameter set.
The overall spectral shape was mostly reproduced by the calculation \cite{SIAM1}, but some discrepancies were found between the experiment and the calculation.
In particular, the peak intensity of $4f^1$ peak was weaker, and its peak position was located at the deeper--photon energy side in the calculation when compared to the experiment. 
To reproduce the XAS spectrum, we took a different parameter set, especially for $E_{\mathrm{F}} - \varepsilon^0_f=-2.3~\mathrm{eV}$ \cite{suppl}.
This problem is presumably caused by the presence of two different $\mathrm{Ce}$ sites in this compound and not considered in the present analysis.
Even in the case of the XAS spectrum parameters, the occupation number of the \Cef states \nf{} was estimated to be 0.766; and thus, \nf{} should be $0.8-0.9$ in both cases.

Previous studies have suggested that the DOS at \EF are dominated by the \Ird states, and \CeIr has been considered a superconductor dominated by the \Ird bands.
However, the present study revealed that the \Cef states contribute significantly to the DOS at \EF and thus to the superconductivity. Therefore, \CeIr is a $s$-wave superconductor with a considerable \Cef contribution as in the case of $\mathrm{CeRu}_2$ \cite{CeRu2_PES}.

Accordingly, all these results suggest that the \Cef states have an itinerant character, and a considerable contribution to the Fermi energy.
The occupation number of the \Cef states \nf{} in the ground state is estimated to be $0.8-0.9$, which was larger than that in the previous studies.

\section{Conclusion}
We studied the electronic structure of \CeIr by the \Cedf resonant photoelectron, $\mathrm{Ce}~3d$ XAS, and $\mathrm{Ce}~3d$ core-level spectroscopies.
All experimental data suggested the well-hybridized nature of the \Cef states with the occupation number of $n_{f} = 0.8-0.9$, which was consistent with the DMFT calculation result \cite{CeIr3_DMFT}.
The comparison of the intensities at the Fermi energy of the on- and off-resonance spectra suggested that more than 30 \% of the DOS at the Fermi energy was contributed by the \Cef states.
These results claimed a finite contribution of the \Cef states on the superconductivity of \CeIr.
Thus, \CeIr is an $s$-wave-type superconductor with a finite contribution from the \Cef states.

\ack
The experiment was performed under Proposal Nos. 2019A3811, 2019B3811, and 2020A3811 at SPring-8 BL23SU.
The present work was financially supported by JSPS KAKENHI Grant Numbers JP16H01084, JP18K03553, JP19J20539, JP20KK0061, and JP	22H03874.

\appendix
\section{Influence of Auger processes in RPES spectra}
\begin{figure}
	\centering 
	\includegraphics[scale=0.5]{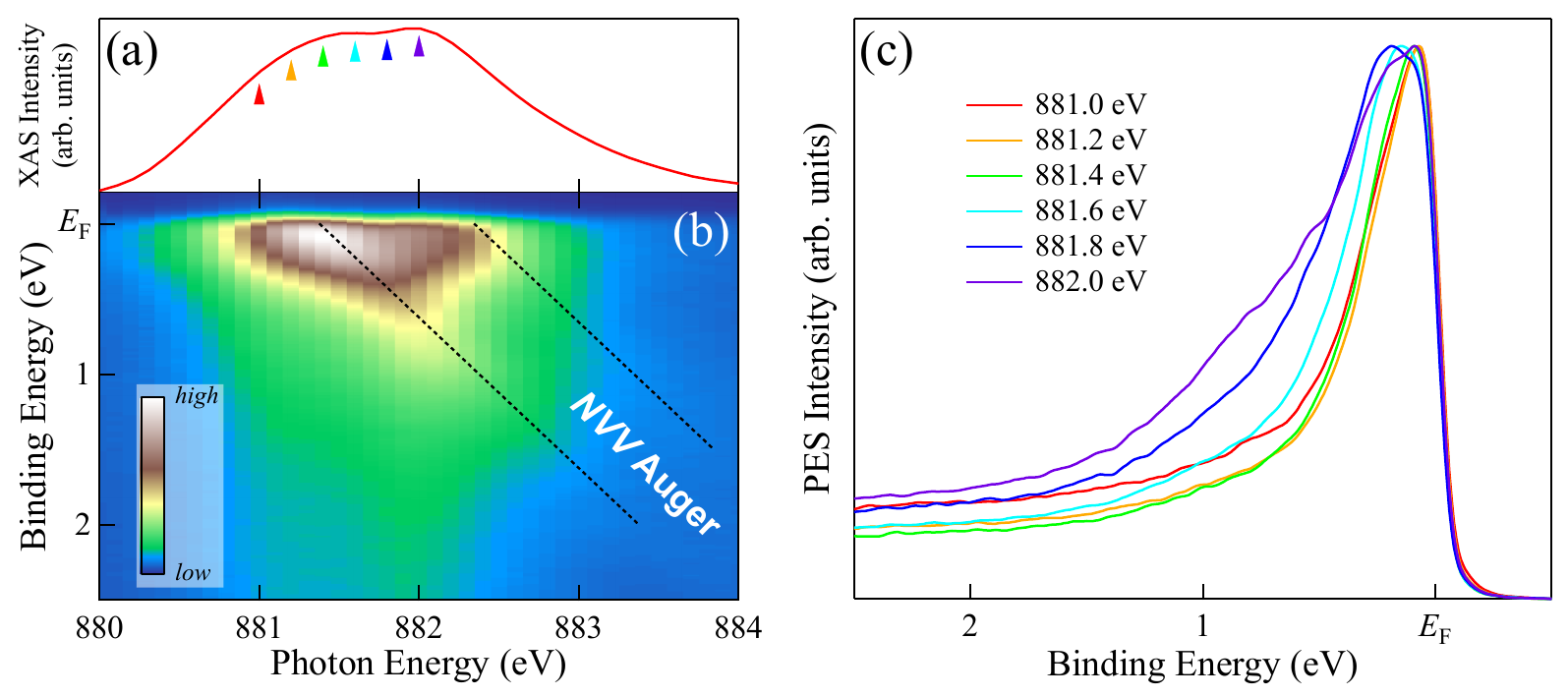}
	\caption{Resonant photoemission spectra of \CeIr.
	(a) Blowup of the $\mathrm{Ce}~4f_{5/2}$ XAS spectrum of \CeIr. 
	(b) Density plot of the RPES spectra of \CeIr as a function of the photon energy.
	(c) Comparison of the RPES spectra of \CeIr, with intensities normalized to their maxima.
	}
	\label{RPES_Auger}
\end{figure}
The contribution from the Auger signals in the resonant photoemission spectrum of the $\mathrm{Ce}$-based compounds was recognized \cite{Ce3d4f_Auger1,Ce3d4f_Auger2,SF_review_JPCM}, and an appropriate selection of excitation energies is required for the measurement of the on-resonant photoemission spectrum.
In this appendix, we describe the influence of Auger signals in the resonant photoemission spectra.
Figures~\ref{RPES_Auger} (a) and (b) show the $\mathrm{Ce}~3d_{5/2}$ XAS and the density plot of the valence band spectra measured at \hn{=880-884}, respectively.
The gigantic resonant enhancement of the \Cef signals at the $\mathrm{Ce}~3d_{5/2}$ absorption edge was recognized at approximately \hn{= 881-882}.
Meanwhile, two diagonal lines were recognized in Fig.~\ref{RPES_Auger}(b), as indicated by the dotted lines, which depict the contributions from the $NVV$ Auger signals.
The resonant photoemission spectra measured at \hn{=881-882} normalized by the peak height of each spectrum is displayed in Fig.~\ref{RPES_Auger}(c)  to examine the impact of the Auger signals on the \Cef spectral profiles.
The spectral shapes of the resonance spectra measured at \hn{=881-881.4} were almost identical, but the higher-binding-energy side of the spectrum was broadened as the photon energy was further increased.
This broadening was the contribution of the Auger signals; thus, the on-resonant spectrum should be measured at \hn{\lesssim 881.4}.
We have used the incident photon energy of \hn{=881.3} for the on-resonance spectrum to avoid the contribution from the Auger signals in the valence-band spectrum.

\section*{References}

\bibliographystyle{iopart-num}
\bibliography{CeIr3}

\end{document}